# Two types of superconducting domes in unconventional superconductors

Tanmoy Das[1], Christos Panagopoulos[2]

[1] *Department of Physics, Indian Institute of Physics, Bangalore 560012, India*
[2] *Division of Physics and Applied Physics, School of Physical and Mathematical Sciences, Nanyang Technological University, Singapore 637371*

Uncovering the origin of unconventional superconductivity is often plagued by the overwhelming material diversity with varying normal and superconducting (SC) properties. In this article, we deliver a comprehensive study of the SC properties and phase diagrams using multiple tunings (such as disorder, pressure or magnetic field in addition to doping and vice versa) across several families of unconventional superconductors, including the copper-oxides, heavy-fermions, organics and the recently discovered iron-pnictides, iron-chalcogenides, and oxybismuthides. We discover that all these families often possess two types of SC domes, with lower and higher superconducting transition temperatures $T_c$, both unconventional but with distinct SC and normal states properties. The lower $T_c$ dome arises with or without a quantum critical point (QCP), and not always associated with a non-Fermi liquid (NFL) background. On the contrary, the higher-$T_c$ dome clearly stems from a NFL or strange metal phase, without an apparent intervening phase transition or a QCP. The two domes appear either fully separated in the phase diagram, or merged into one, or arise independently owing to their respective normal state characteristics. Our findings suggest that a QCP-related mechanism is an unlikely scenario for the NFL phase in these materials, and thereby narrows the possibility towards short-range fluctuations of various degrees of freedom in the momentum and frequency space. We also find that NFL physics may be a generic route to higher-$T_c$ superconductivity.

## CONTENTS

# I. INTRODUCTION

Unconventional superconductivity has been observed in a growing class of materials, including heavy fermions,[1] copper-oxides,[2] organic salts,[3] and more recently in iron-pnictides,[4] iron-chalcogenides,[5] and oxybismuthides[6]. Despite more than three decades of research, a concluding mechanism for the electron pairing in these diverse families of superconductors has not prevailed[7]–[12]. The challenge is limited not only to understanding the superconducting (SC) state, but also their corresponding anomalous normal state features, which differ from one material to another and thereby hinder the extraction of the fundamental, and presumably common parameter(s) responsible for unconventional pairing.

In these systems, superconductivity often arises when the material is driven away from its pristine composition and / or ambient condition, and yields a SC dome. It is often seen that the optimum SC transition temperature $T_c$ is possibly neared by a quantum critical point (QCP) of an intertwined electronic order and / or linear-in-temperature resistivity, attributed to a non-Fermi liquid (NFL) behavior[13], [8], [9], [14], [15] However, so far no universal link has been established between the SC dome, a QCP and / or the NFL state.

The presence of SC domes is known in conventional superconductors since 1965 in doped $SrTiO_3$[16], subsequently in Li,[17] Fe metal,[18] and more recently in gated $LaAlO_3$/$SrTiO_3$ interface,[19] two dimensional $MoS_2$,[20] as well as in other transition metal dichalcogenides[21]. Similarly, the NFL physics has been observed in a larger class of materials, beyond unconventional superconductors, extending to the less correlated semiconductors,[22] disordered systems[23], and the skyrmion system MnSi.[24] Furthermore, there is an increasing number of unconventional superconductors in which the NFL state is far from a QCP of an order phase, or arises inside the order phase, or even occurs in absence of a QCP.[25]

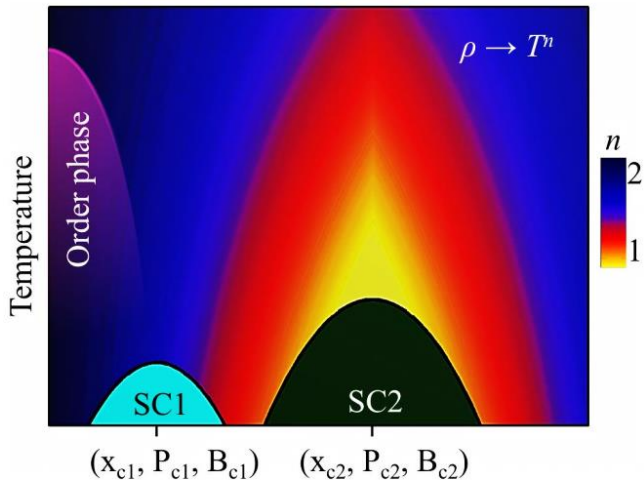

**Fig. 1. The phase diagram (schematic) of two SC domes in unconventional superconductors.** This phase diagram is qualitatively similar, at low temperature, in a large class of materials that belong to all families of unconventional superconductors. In materials where the two SC domes are isolated, superconductivity emerges at disconnected locations of the ground state tuning parameters. (Here, $x_{ci}$, $P_{ci}$ and $B_{ci}$ are the characteristic values of doping, pressure, and magnetic field, respectively, where superconductivity is optimum.) The background color-gradient depicts the resistivity-temperature exponent ($n$), indicating that the normal state of the second SC dome is a NFL state. The electronic/magnetic order and its QCP are materials dependent and sometimes absent.

A viable protocol to clarify the role of a QCP, NFL physics and their possible role in higher-$T_c$ superconductivity may be via individual tuning of these properties through the application of multiple control parameters such as disorder, pressure, magnetic field on a doped sample, and vice versa. For example, disorder quenches unconventional superconductivity,[26], [27] which in turn helps unveil the location of a QCP with respect to a SC dome.[28]–[31] Systematic studies in cuprates have demonstrated that disorder can fully suppress superconductivity at the 1/8 doping, leaving two separated SC domes. However, the QCP of the parent antiferromagnetic (AFM) phase often remains unperturbed and experiments suggest that it coincides with the lower SC dome.[28], [32] On the other hand, the linear-in-$T$ dependence of the resistivity, or the NFL physics continue to dominate at the second SC dome region at all disorder levels.[33] Pressure provides another universal route to expose the two SC domes structure in several families of superconductors including cuprates,[34] heavy-fermions,[30], [35], [36] iron-pnictides,[37], [38] iron-selenides,[39] as well as organic superconductors[40]. Similarly, magnetic field is another effective parameter to reveal the two domes features in cuprates [41] and heavy-fermions[42], [43].

In this article, we report a holistic analysis of a gamut of experimental data across all the above mentioned families of unconventional superconductors and discuss the emergence of an apparently generic phase diagram of superconductivity, as illustrated in Fig. 1. Analyzing an extensive range of literature, we find that the SC dome nearer a possible QCP (SC1) has lower $T_c$ than the SC dome (SC2) surrounding a NFL region (Fig. 1). Both domes are unconventional, but with characteristic SC and normal state properties. The lower-$T_c$ dome is present with or without a QCP, but in the absence of a NFL state. On the other hand, we find the higher-$T_c$ SC dome consistently arises from a NFL state, seemingly in the absence of a QCP, in all families. Furthermore, we find that higher $T_c$ emerges where the normal state (residual) resistivity is lower and linear in temperature. Detailed analysis of an extensive set of data obtained using thermodynamic, transport and spectroscopic experiments demonstrates distinct associated anomalies in the two SC domes, characteristic of their pairing states. We discuss briefly theoretical proposals that have been put forward for the NFL state and comment on the challenges emerging from our findings on the fundamental role of NFL towards high-$T_c$ superconductivity. Finally, we summarize with tentative guidelines for future research employing multiple tuning and a possible route towards higher-$T_c$ superconductivity. In the appendix, we present a list of materials, separating those with apparently one SC dome (where QCP and NFL state coincide, or either a QCP or NFL state is present), and those showing two domes, where the QCP and NFL states are well separated. Clearly, the number of the latter materials is considerably larger than the former.

## II. HEAVY FERMIONS

We start with the older class of unconventional superconductors namely, heavy fermions (HF's). These materials exhibit a rich variety of correlated phenomena including Kondo physics, valence fluctuations, magnetic (AFM, or ferromagnetic (FM) in different materials) QCP, NFL state, and unconventional superconductivity[7], [9]. CeIn$_3$, CePd$_2$Si$_2$, YbRh$_2$Si$_2$ and UCoGe are among the leading HF's where a NFL state seemingly coincides with the QCP of the ordered phase, and in some cases with optimum $T_c$ as well. However, there is a growing class of HF materials[25], [44], [45] where the NFL physics emerges either without a QCP or far from the QCP, or even sometimes within the ordered phase. A list of HF materials with NFL behavior in the absence of a QCP can be found in Ref. [25]. A leading example is β-YbAlB$_4$ where a NFL feature appears at a small SC dome without tuning the material to a QCP[46]. Similarly, CeCoIn$_5$ is considered a 'born' NFL superconductor without QCP; here a QCP can be engineered via selective chemical doping or the application of a magnetic field. However, as the QCP is introduced, the NFL state is unexpectedly reduced.[47]–[50] Furthermore, in UGe$_2$,[51] and ZrZn$_2$[48] a SC dome is present only inside the FM order phase and is suppressed with pressure at the FM QCP.

### IIA. The CeCu$_2$Si$_2$ family: Dissecting QCP and NFL

CeCu$_2$Si$_2$ is one of the earlier materials where the splitting of a SC dome in two separate domes at the disconnected QCP and NFL points was observed [30] (Fig. 2(a)). In the pristine phase, the system exhibits an unusual SC dome, where optimal doping is substantially departed from the AFM QCP as a function of pressure. Ge-doping leaves the AFM phase almost unchanged,[53] while superconductivity is suppressed and splits into two isolated SC domes. The small SC dome centers around the AFM critical point, whereas the large dome persists to a higher pressure. The resistivity power-law exponent $n$, defined by $\Delta\rho(T) = [\rho(T) - \rho(0)] \propto T^n$, is higher at the AFM QCP ($n \geq 1.5$) and depends strongly on Ge doping, despite the AFM order being robust to it [53]. As in cuprates, $n \to 1$ at the second SC dome and remains fairly doping / disorder independent. This suggests that the NFL state stems from small-angle scattering, in agreement with the density- and/or the valence fluctuation approach [9], [53]. In the fully doped sample CeCu$_2$Ge$_2$, superconductivity reemerges at the AFM QCP [54]. It would be interesting to study the systematic doping dependence for the entire doping range of CeCu$_2$(Si$_{1-x}$Ge$_x$)$_2$. In the related compound CeRu$_2$(Si$_{1-x}$Ge$_x$)$_2$, no enhancement of the effective mass is detected by ARPES across the AFM QCP, although it is expected to diverge here.[55]

In fact, a strain engineering study in the related superconductor CePd$_2$Si$_2$ showed that all three features namely, AFM QCP, NFL and SC dome acquire the same strain dependence [56].

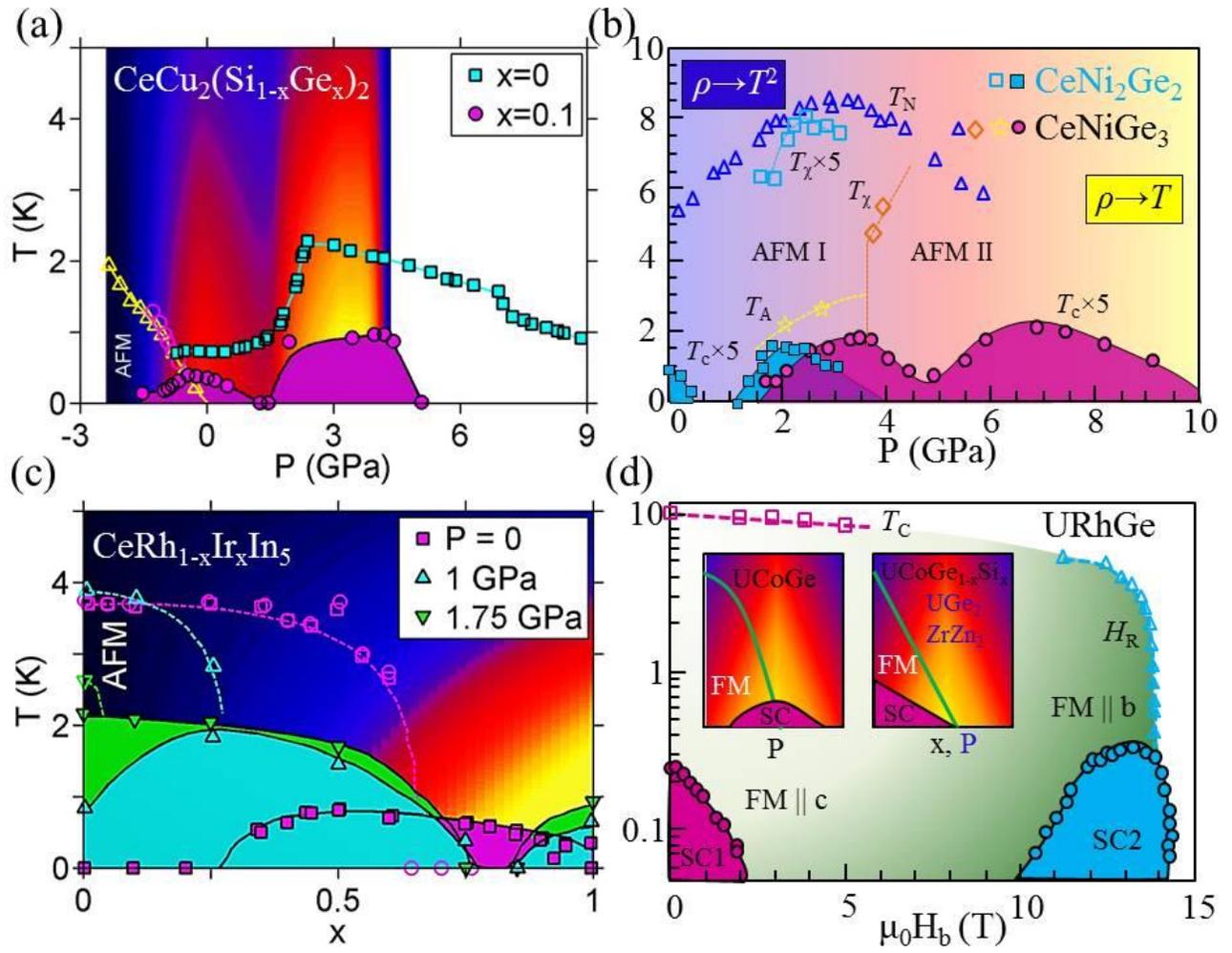

Fig. 2. Two SC domes in various heavy fermion compounds under different conditions. (a) Pressure tuned phase diagram of $CeCu_2(Si_{1-x}Ge_x)_2$ at two dopings.[30], [53] $T_N$ data is presented for x = 0.1 (magenta) and x = 0.25 (yellow). The background color-map depicts the same resistivity exponent as in Fig. 1. (b) Similar two domes phase diagram for $CeNi_2Ge_2$ from Ref. [57] and $CeNiGe_3$ from Refs. [58] $T_N$, $T_x$ and $T_c$ are deduced from a kink in $\rho(T)$, peak in $d\rho(T)/dT$, and the onset of $\rho = 0$, respectively. $T_A$ marks the temperature where incommensurate to commensurate crossover occurs in the NQR data. (c) Two SC domes revealed by pressure in Ir-doped $CeRhIn_5$.[35] The resistivity exponent, $n$, shown in color map is interpolated from three values at x = 0, 0.5 and 1 for $P = 0$. (d) Magnetic field dependent phase diagram of URhGe having two SC domes.[42], [43] $T_C$ is the Currie temperature and $H_R$ the crossover field where the magnetization easy axis changes from the $c$- to $b$-axis [43]. The white to green color-map depicts the strength of the $b$-axis magnetization [43]. Two insets show the contrasting phase diagram of a similar UCoGe system having a SC dome at the FM QCP. Si doped UCoGe as a function of doping,[59] and $UGe_2$ [47] and $ZrZn_2$ [52] as a function of pressure depict both SC and FM orders to disappear together.

$CeNi_2Ge_2$ and $CeNiGe_3$ exhibit a similar two domes phase diagram as a function of pressure[57], [58] (Fig. 2(b)). The Néel temperature in this system acquires a dome feature. Inside this dome we have the presence of two distinct SC domes. Their optimum pressures coincide with maximum $T_N$ and the QCP of $T_N$, respectively. Furthermore, the upper critical field $H_{c2}$ shows similar pressure dependence and the resistivity exponent $n$ decreases with pressure, suggesting that the second SC dome is NFL-like. Subsequently, a hidden QCP was identified between two AFM phases at the optimum value of the first SC dome, where a transition from commensurate to incommensurate AFM order occurs. The commensurate AFM-I phase coexists with superconductivity while the incommensurate AFM-II order is competing, suppressing superconductivity in between the two SC domes. Notably, this scenario is reminiscent of the 1/8 doping in cuprates, as discussed in Fig. 3, where incommensurability is dominant precisely where superconductivity is suppressed, raising the possibility that the two SC domes are caused by homogeneous and inhomogeneous normal states, respectively. This picture however, loses support from other materials classes.

**IIB. The CeCoIn$_5$ family: 'Born' NFL systems**

Ce$M$In$_5$ ($M$ = Co, Ir and Rh) is a widely studied HF class in which SC, QCP and NFL phases can be monitored separately. CeCoIn$_5$ shows superconductivity without tuning and is a 'born' NFL system without an apparent QCP. CeIrIn$_5$ is neither magnetic nor SC in its pristine phase. With Co-doping on the Ir site, superconductivity develops uniformly with doping, reaching maximum in the fully doped CeCoIn$_5$, without magnetic order. On the other hand, CeRhIn$_5$ is an AFM metal with a resistivity exponent close to or above 2, hence the system is away from the NFL state. It becomes a superconductor at $P \sim 1.6$ GPa with $T_c \sim 2$ K.[60] AFM order can be introduced to these samples by Rh doping at the $M$ site,[47] or Cd/Hg doping at the In site [49], [50]. In this way, an AFM QCP is introduced to the sample. Interestingly, no enhancement or anomaly in $T_c$, or in $n$ takes place as the material passes though the QCP [48], see Appendix A.

A pressure dependent study on CeRh$_{1-x}$Ir$_x$In$_5$ shows how the single SC dome at the AFM QCP splits into two [35] (Fig. 2(c)). Here, the first dome at the QCP acquires a higher $T_c$ and both the QCP and optimum $T_c$ shift simultaneously to lower doping with increasing pressure. The deduced values for $n$ consistently indicate that the dominant NFL doping departs from the QCP and the second SC dome occurs where $n$ is lowest. On the other hand, $n$ is strongly pressure dependent, decreasing from ~2 to ~1 at $x = 0$ with increasing pressure, while it remains NFL-like ($n \leq 1$) at $x = 1$. A complete phase diagram of multiple tuning at various atomic sites for Ce$_x$R$_{1-x}$M(In$_{1-y}$T$_y$)$_5$ [$R$=Yb, La; $T$=Hg, Cd, Sn, Pt] is given in Appendix A, providing further insight into the nature of QCP, NFL and superconductivity in this family.

**IIC. The UCoGe family: FM and possible 'triplet' superconductivity**

Coexistence and competition of FM order with superconductivity was discovered in several uranium and zirconium based compounds, with contrasting phase diagrams. In UCoGe, the SC dome center coincides with a FM QCP as a function of pressure,[61] while superconductivity is present only within the FM order state and vanishes at the same pressure with ferromagnetism in UGe$_2$,[51] ZrZn$_2$ [52] (see insets in Fig. 2(d)). Si doping in UCoGe$_{1-x}$Si$_x$ strongly suppresses superconductivity,[59] and confines the latter within the FM order state, as seen in the pressure tuned systems UGe$_2$ and ZrZn$_2$. A similar FM and SC phase diagram emerges with applied magnetic field along the $b$-axis, perpendicular to the easy magnetization axis [62]. On the other hand, Fe and Ni dopings on the Co site suppress superconductivity by about 4% doping, while ferromagnetism persists above 20% doping[63]. Taken together, the study of multiple tuning suggests that the SC dome may not be universally associated with the FM QCP in UCoGe.

The magnetic field ($H//b$-axis) study in isostructural URhGe reveals two isolated SC domes, which lie about 8 T apart.[42] Spontaneous magnetization has the easy axis parallel to the $c$-axis and with applied magnetic field along the $b$-axis the spins are gradually tilted towards the field direction [the magnetization values along the $b$-axis is shown by green shading in Fig. 2(d)]. With a smaller field of 2 T the first SC dome is fully suppressed. The total magnetization remains field independent up to 16 or 17 T. Furthermore, for B ~ 10 - 11 T the magnetic moment is fully tilted from the $c$-axis to the $b$-axis. Simultaneously, a second SC dome emerges around the same field strength with a dome-like feature persisting up to ~ 14 T. $H_R$ depicts the temperature dependent critical field for which the easy magnetization switches from $c$- to $b$-axis [43], indicating that the second SC dome is linked to the in-plane magnetization, whereas the first dome to its $c$-axis component. No compelling evidence for a QCP and a NFL behavior is reported in these studies.

M. B. Maple discussed a wide range of materials, where the NFL state arises without an easily detectable QCP, or inside the ordered phase, or far from the QCP. For example, in the U$_{1-x}$M$_x$Pd$_2$Al$_3$ ($M$ = Th, Y, La) family, the SC state appears only inside the AFM state, and disappears earlier than the AFM QCP.[64] Interestingly, the NFL state arises above the QCP and is often intervened by a spin-glass (SG) phase. $M_{1-x}$U$_x$Pd$_3$ ($M$ = Sc, Y) has a similar phase diagram without a SC phase.[65], [66] In Yb$_2$Fe$_{12}$P$_7$, the NFL phase is observed deep inside the magnetic phase, not at the QCP,[67] which is in sharp contrast to the YbRh$_2$Si$_2$ phase diagram.[68] Similarly, in URu$_{2-x}$Re$_x$Si$_2$, NFL is also found inside the FM phase.[69]

# III. CUPRATES

Cuprates exhibit high-$T_c$ superconductivity when doped away from the Mott insulating phase. As shown in Fig. 3, in hole-doped cuprates, the SC dome has a universal 1/8 doping anomaly where superconductivity is suppressed - the suppression of $T_c$ is material dependent. Here, we present results for various members of the cuprate family and show that through the utilization of distinct tuning parameters such as disorder, magnetic field, or pressure, cuprates give rise to two isolated, albeit intrinsic SC domes.

## IIIA. Disorder induced two domes

Among the high-$T_c$ cuprates, $La_{2-x}Ba_xCuO_4$ (LBCO) is known to possess maximum intrinsic disorder, which is reduced gradually in $La_{2-x}Sr_xCuO_4$ (LSCO), $Bi_2Sr_2CaCu_2O_{8+x}$ (Bi2212), $YBa_2Cu_3O_{7-y}$ (YBCO), and $HgBa_2Ca_4Cu_5O_{12+\delta}$ (Hg-1245). The SC $T_c$ can be tuned by impurity substitution, as amply demonstrated for example, by Koike *et al.*[28], [29], [32] Alloul *et al.* [27], [70], Bianconi and Missori [71], and Ando *et al.* [33]. With the substitution of Cu with Ni, Zn and Ga in LBCO [28] and LSCO [29], the splitting of the SC envelop into two domes becomes apparent [Figs. 3(a) and 3(b)]. Interestingly, for Ga doping, the optimum $T_c$ of the second SC dome is lower than the first one. Similarly two SC domes can be deduced in Co and Zn-doped YBCO [72]. Alloul *et al.*[27], [70] have shown that with sufficient Zn substitution in YBCO, the SC dome in the lower-doping regime is suppressed while the AFM state shifts to lower doping, allowing only the second dome to persist. These results suggest that the details of the underlying pairing states may be different in the two SC domes, responding differently to different impurities.

Various multilayer cuprates have been grown to date, mainly in the Bi-, Y-, Tl-, and Hg-families. These systems give additional flexibility to study the SC phase diagram. Studies on different multilayer systems have demonstrated the intervention of the AFM order (either long-range or short-range) into the SC dome and the critical doping for the AFM order is found to be layer dependent[73]–[76]. The phase diagram of *M*-1245 (*M*=Hg, Tl, Cu) [75] is indicative of the presence of two SC domes, which can be revealed by multiple tuning. The resistivity of Hg-cuprates[77] is also linear-in-*T* at optimal doping and quadratic for x ~ 0.1.

The intervening normal state and QCP physics are largely debated in cuprates due to the presence of several intertwined phases including AFM, SG, pseudogap, charge density wave (CDW), nematic state, and stripes, as we discuss in Sec. VIB. Stripes have been widely studied, especially due to the dominance at the 1/8 doping where superconductivity is suppressed in hole-doped cuprates. The strength of stripe order is highest in LBCO and is gradually reduced in LSCO, Bi2212, YBCO and Hg-based cuprates. Recently, CDW is observed in YBCO, Bi1201, Bi2212 around the 1/8 doping. Contradictory evidence of a QCP in YBCO at the second dome is also presented. While in Y123 compound,[79] the effective mass appears to diverge, in Y124 it rather reduces at optimal doping,[80] indicating the absence of a QCP. Notably, these electronically complex phases, present in some members of the hole-doped cuprate family, are not necessarily present in other families of unconventional superconductors. Hence, the role of a cuprates-specific CDW, stripes, and a pseudogap state is not discussed in detail here. Instead, we focus on electronic phases and properties for which experimental evidence over the past three or four decades, indicate as common across the families of unconventional superconductors. In general, in the underdoped regime of hole-doped cuprates the AFM order decreases sharply and near its end point a SG state and/or a short-range spin-density wave (SDW) develop. The doping where a possible QCP of the SG/SDW order phase occurs [Fig. 3(a)] coincides with the optimum doping of the first SC dome. In some QCP systems, the leading exponent *n* reduces to ≤ 1 in reaching the QCP.[9] On the other hand, in the case of cuprates, *n* exhibits no considerable anomaly at this apparent SG/SDW critical doping. Instead, *n* decreases gradually, reaching a minimum value (≤1) at the optimal doping of the second (higher $T_c$) SC dome. While there have been suggestions for the possible presence of a QCP at the second SC dome, based on measurements other than the resistivity data [79], [81]–[83], this scenario requires further investigation. Although there is no clear evidence for an actual order extending to the second SC dome, ample evidence indicate that short-range magnetic correlations extend to the second SC dome and up to optimal doping i.e., where the absolute value of the superfluid density is maximum[83].

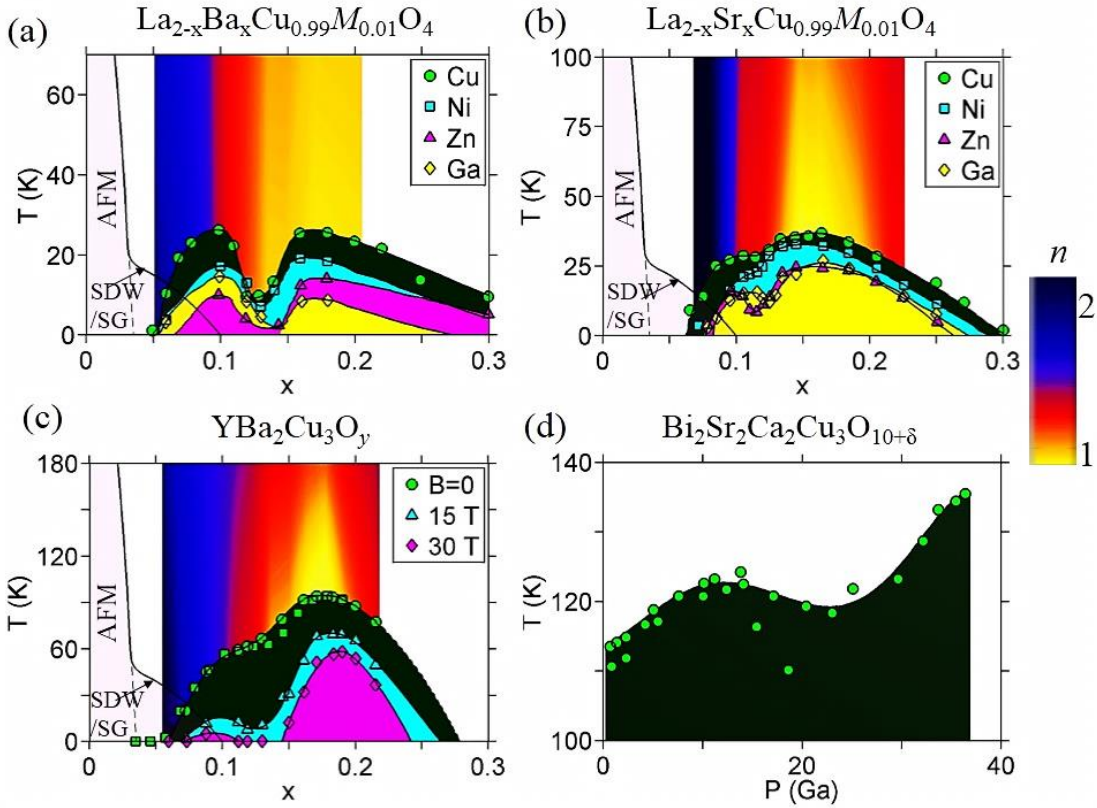

Fig. 3. Evolution of phase diagrams in various cuprates with multiple tunings. The color plot in the background depicts the variation of the resistivity exponent ($n$), for the same material, but without disorder (a,b) and at zero magnetic field (c). (a-b) LBCO data are taken from Refs. [28], [32]. LSCO data from Refs. [29], [33], [78]. (c) Magnetic field dependent data for YBCO are taken from Ref. [41]. The resistivity exponent of YBCO is extracted from the data in Ref. [33]. (d) The pressure dependent data in $Bi_2Sr_2Ca_2Cu_3O_{10+x}$ (Bi2223) are obtained from Raman scattering in Ref. [34], for which the corresponding resistivity data is unavailable. The magnetic phase consists of the Néel order at high temperature and a SG or a SDW near $x = 0.1$ (see Fig. 3(a)).

There is also evidence for two SC domes in some electron doped systems, such as $Sr_{1-x}La_xCuO_2$[84] and $La_{2-x}Ce_xCuO_4$ (LCCO)[85]. The long tail in $T_c$ above the optimal $T_c$ in LCCO is reminiscent of the SC dome seen in La-1111, or the hole doped 122 pnictide families, and hints to the presence of a second dome which can be revealed by multiple tuning. Although $T_c$ in both SC domes is strongly disorder dependent, experiments reported that the QCP and the NFL state are often less sensitive to disorder [28]. This again suggests that the presence of two SC domes is not an artifact due to impurities but an intrinsic property resulting from the separation between the QCP and the NFL state - revealed when superconductivity is quenched by disorder.

### IIIB. Magnetic field

Taillefer and co-workers [41], [86] have further demonstrated the presence of two SC domes in YBCO by the application of an external magnetic field (Fig. 3(c)). A field of ~30 T is sufficient to reveal two disconnected SC domes. Notably, the optimal dopings in both domes are insensitive to the applied field strength, a trend similar to the case of impurity substitution (Fig. 3(a-b)). Finally, the resistivity exponent $n$ at zero field (extracted from Ref. [33]), concurrently, reveals the same doping dependence as discussed before, in that the second SC dome acquires optimum $T_c$ when $n$ is minimum ($\leq 1$).

### IIIC. Pressure

A similar two domes feature is also established as a function of pressure in Bi2223 near optimal doping,[34] (Fig. 3(d)). The pressure dependent study is performed through Raman scattering and thus the pattern of the associated QCP and NFL is unknown along this parameter axis.

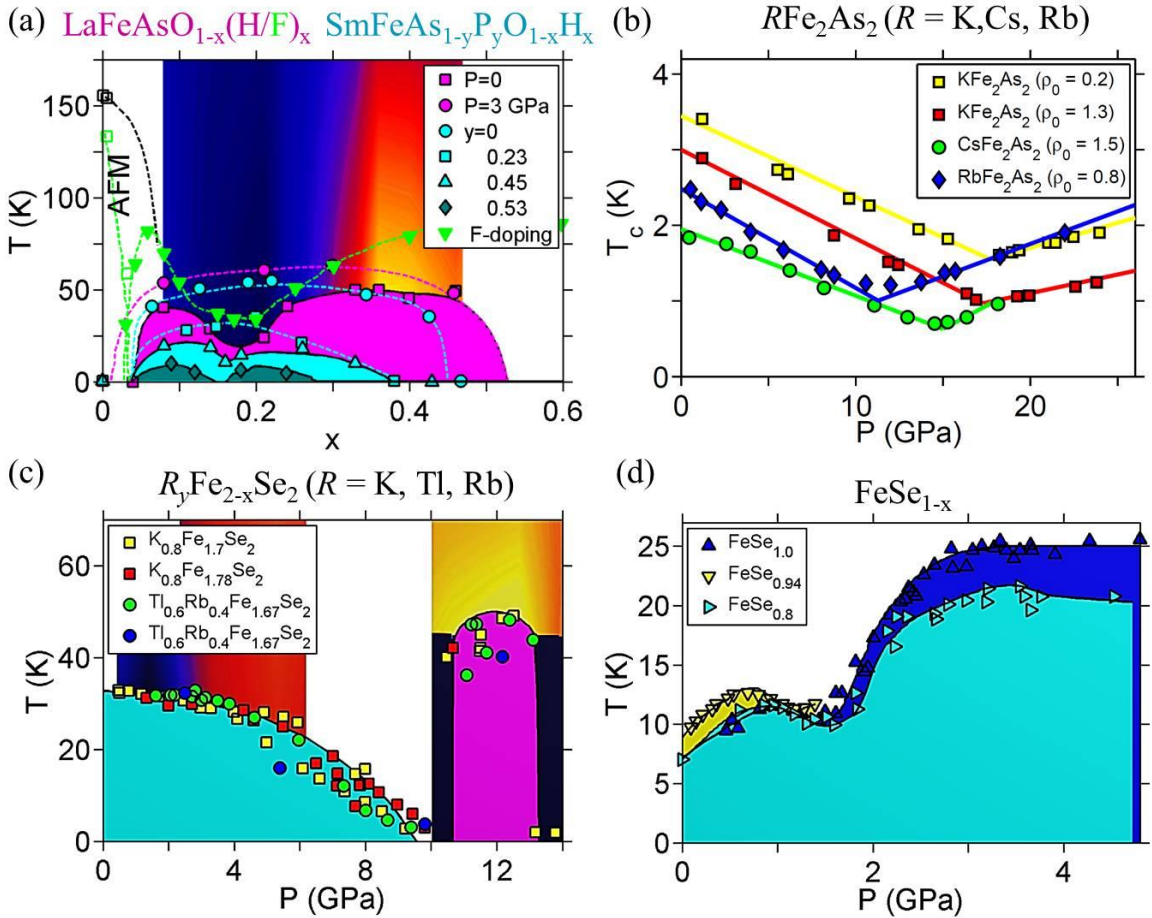

**Fig. 4.** Multiple SC domes in iron-pnictides and chalcogenides. (a) Hydrogen substitution on the oxygen site in the '1111' oxypnictides gives rise to two SC domes.[31], [37], [87] The two domes structure is revealed in La-1111 through doping, while in Sm-111 is achieved by P substitution on the As site. The resistivity exponent is shown in color plot for La-1111 at $P = 0$ [37]. F-doped La-1111 is also shown recently to inherit two SC domes (green symbols).[88] (b) Pressure tuned two SC phases in a 122 family which has only hole-pocket FS.[38], [89], [90] The high-pressure SC phase is destroyed with slight electron doping by Co substitution. (c) Two fully isolated SC domes are observed in '122' based iron-chalcogenides under pressure.[39] The resistivity exponent is plotted for only one sample. (d) Two SC domes are also reported in the '11' family.[91], [92]

## IV. PNICTIDES AND CHALCOGENIDES

Soon after the discovery of unconventional superconductivity in LaFeAsO$_{1-x}$F$_x$ in 2008,[4] a variety of SC materials within the iron-arsenic and iron-selenide families have been synthesized.[10] In these materials, doping suppresses the SDW order giving way to a SC dome.[14], [15] However, in several compounds either within the same or in different families, superconductivity occurs in the absence of a magnetic order, or far from a QCP [39], [89], [93]. Despite most of these materials exhibiting a nodeless SC gap, the underlying pairing mechanism is believed to be unconventional [11]. We will show that like in the HF and cuprate families, here too there is increasingly systematic evidence for the presence of two types of SC domes.

### IVA. The 1111 family

The fluorine doped '1111' family – $R$FeAsO$_{1-x}$F$_x$ ($R$ = La, Ce, Sm, Gd) – shows contrasting SC phase diagrams in different materials.[94], [95] While in La-1111, an abrupt (first-order-like) transition occurs from a SDW state to superconductivity with F doping,[95] the same doping in Ce-1111 gives rise to a second order phase transition - from the complete elimination of SDW to the emergence of superconductivity[94]. In Sm-1111 however, SDW and superconductivity coexist [96]. A two-SC-domes phase diagram is reported in LaFeAsO$_{1-x}$F$_x$ with a QCP at the first dome and a NFL at the second dome in the absence of a QCP (green symbols in Fig. 4(a)).[88] Hosono *et al.* have recently doped 1111 oxypnictides with hydrogen, introducing relatively large disorder to the sample [31], [37], [87]. LaFeAsO$_{1-x}$H$_x$ exhibits two SC domes, with the first dome emerging at the SDW QCP whereas the second dome occurs at the peak of the NFL feature [37], see Fig. 4(a). (In the other three rare-earth elements (Ce, Sm, Gd), an apparent single but wide SC dome is known so far in the absence of multiple tunings.) The resistivity exponent $n$ is NFL-like at all H dopings except in the La-compound. In the latter family, a transition from NFL to FL ($n \geq 2$) occurs with underdoping, coinciding with the second and first SC domes, respectively. A later study on H-doped Sm-1111 with additional isovalent substitution of P on the As site, observed that the single SC dome gradually splits into two separate domes Fig. 4(a) [31]. Moreover, their analysis reveals that Sm-1111 is NFL-like at all H-dopings without P-substitution; then exhibits a H-doping dependent NFL state as P content is increased; and finally crosses over to a FL-like behavior with further increase of P substitution. LaFeAs$_{1-x}$P$_x$O is one of the few systems where the presence of a QCP beneath the SC dome has been revealed from a number of experimental observations. These include studies of the resistivity exponent,[97] NMR,[98] and magnetic penetration data[99]. With further doping, a second AFM state emerges[100]. At higher doping, the second AFM phase again disappears and towards its end point a second SC dome emerges. Here the QCP is presumably far below the optimum doping of the SC phase. Superconductivity persists up to the fully doped LaFePO, which is actually the first discovered member of the pnictide superconductor family. F-doping on the O site gives rise to a SC dome without an apparent magnetic order. The enhancement of superconductivity in the second dome at a higher doping in LaFeAs$_{1-x}$P$_x$O and LaFePO$_{1-y}$F$_y$ is argued to stem from the topological Fermi surface (FS) transition.[100] No result for the temperature dependence of the resistivity is available to study the possible reappearance of a NFL state in the two SC domes. Considering the similarity to H- and F-doped iron-arsenides, we envision that multiple SC domes may be hidden in this compound. Multiple tuning can be employed to study this possibility.

**IVB. Fully electron and hole doped 122 families**

BaFe$_2$As$_2$ ('122' family) is an extensively studied bilayer superconductor family, which can be doped at all ionic sites to give rise to superconductivity, often nearing a SDW QCP[15]. Hole doped Ba$_{1-x}$K$_x$Fe$_2$As$_2$ shows a wider SC dome whose optimum doping is about 20% higher than the SDW QCP. Electron doping is achieved by Co and Ni substitutions on the Fe site and the SDW QCP and SC dome seemingly coincide at this doping. However, a recent study on Ba(Fe$_{1-x}$Co$_x$)$_2$As$_2$ indicates the presence of a nematic QCP at the optimal doping in this compound.[101] Finally, isovalent substitution of P for As and Ru for Fe, which is equivalent to pressure, gives rise to a SC dome at the SDW QCP [97], [102]. Notably, BaFe$_2$(As$_{1-x}$P$_x$)$_2$ is a unique system where the resistivity exponent and the superfluid density indicate the presence of a NFL feature at the QCP where superconductivity is optimum[15], [103]. Recently, a strain dependence study on thin films of the parent BaFe$_2$As$_2$ (which is not a superconductor at zero strain) shows the emergence of bulk superconductivity without an apparent SDW order or a QCP [104].

Superconductivity re-emerges in the highly doped (hole and electron) members of the 122 family in the absence of an apparent QCP. $M$Fe$_2$As$_2$ ($M$ = K, Cs, and Rb) is a fully hole doped analog of BaFe$_2$As$_2$, in which superconductivity emerges in the pristine material. Interestingly, with pressure superconductivity is suppressed first, and above a critical pressure it starts to remerge,[38] indicating the presence of two underlying pairing states (see Fig. 4(b)). Slight Co doping to the Fe-site suppresses the high-pressure SC phase, but not the lower-pressure phase, suggesting that the two SC phases may be microscopically distinct[105]. On the other hand, superconductivity is monotonically suppressed with applied magnetic field [106]. Resistivity results in KFe$_2$As$_2$ show that this system is a NFL with exponent $n$ ~1.5 without a QCP at ambient condition. It is also found that $n$ increases as $T_c$ decreases with applied magnetic field [106].

A fully electron-doped analog of the BaFe$_2$As$_2$ superconductor is realized in the new class of 122 iron-chalcogenide KFe$_2$Se$_2$ family, which becomes SC when sufficient Fe vacancies are introduced[107]. Clearly, these materials contain disordered vacancies, hence the SC volume fraction is low. Sun *et al.*[39] studied several materials of this family via pressure at their optimal vacancy concentrations. Here superconductivity is first monotonically suppressed and following a non-SC region for approximately 1-2 GPa, a second SC dome emerges with a $T_c$ higher by 10-20 K (Fig. 4(c)). There is no compelling evidence for a magnetic order or a QCP in the SC domes. However, the *T*-dependence of the resistivity indicates $n \leq 1$ in the second SC dome and $n = 1.5 - 2$ in the first one.

The nickel-based 122 family of superconductors was discovered in BaNi$_2$P$_2$[108] the year iron-based superconductors were first synthesized. Subsequently BaNi$_2$As$_2$,[109] SrNi$_2$P$_2$[110] and SrNi$_2$As$_2$,[111] and finally the chalcogenide family TlNi$_2$Se$_2$[112] were discovered. Because the maximum $T_c$ reported to date is less than 5 K, this family had not gained as much attention. Interestingly, these materials are 'born' superconductors without tuning, however, no extensive study to understand the presence of a NFL state has been reported, except the fact that the effective mass of these materials is relatively high ($m = 14m_b$ - $20m_b$)[112]. In a more recent study, multiple tuning with pressure in doped TlNi$_2$Se$_{2-x}$S$_x$ revealed two SC domes[113]. In TlNi$_2$Se$_2$, superconductivity decreases smoothly with pressure, whereas in TlNi$_2$SeS, superconductivity is first enhanced and then suppressed to form a SC dome. Finally, in TlNi$_2$S$_2$ two SC domes are revealed by pressure. The $T_c$ versus doping phase diagram may be an overlap of two SC domes. The mechanism for the two SC domes in this family has yet to be investigated.

### IVC. The FeSe$_{1-x}$ ('11') family

The '11' family of FeSe$_x$Te$_{1-x}$ shows a phase diagram reminiscent of cuprates: the SDW state disappears rapidly with doping and the emergence of superconductivity is intervened by a short-range SG phase.[114] In the parent compound, two connected SC domes are reported as a function of pressure for different Se compositions [91], [92] as reproduced in Fig. 4(d). A magnetic phase is also reported [92] in the second SC region, but the complete magnetic structure is not yet known. The resistivity data is not presented in the pressure dependence study, however, its temperature dependence for α-FeSe$_{0.88}$ [5] and β-Fe$_{1.01}$Se [115] indicates that these are NFL systems.

## V. OTHER MATERIALS

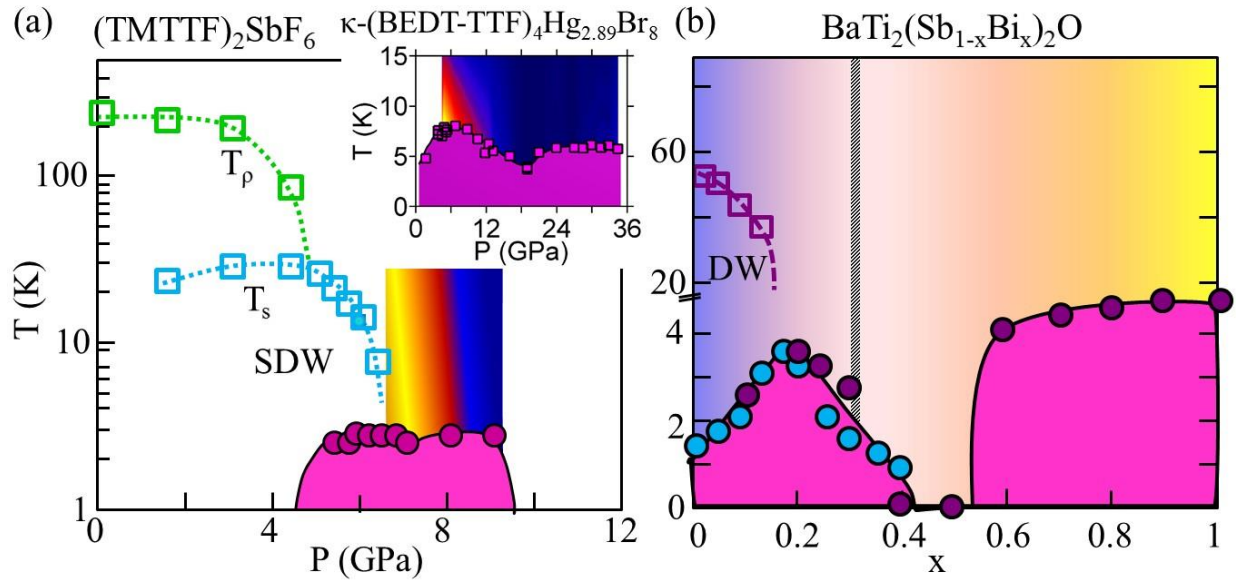

**Fig. 5**: Two SC domes in organic and oxybismuthide superconductors. (a) Two SC domes emerge in two distinct families of organic superconductors with [116], [117] and without a QCP (inset) [40]. A NFL feature is deduced from the power-law dependence of resistivity. $T_s$ and $T_\rho$ are two energy scales where dips in the resistivity are obtained.[116] (b) Phase diagram of Sb doped oxybismuthides having the first SC dome at the QCP of a DW (either SDW or CDW),[118] and the second dome at the NFL side.[119] Comprehensive resistivity studies are not available to deduce the exponent *n* at all dopings.

## VA. Organic superconductors

Superconductivity was discovered in the organic salt (TMTSF)$_2$PF$_6$ at a pressure of 0.65 GPa with $T_c \sim$ 1 K in 1980 [3]. Subsequently, several organic materials classes have been discovered, for example, (TMTSF)$_2$X; (TMTTF)$_2$X; ζ-(BEDT-TTF)$_2$X (ζ = α, β, κ) [117] with highest $T_c$ reaching around 4 K. In agreement with cuprates, heavy-fermions, and pnictides, superconductivity in organic families emerges via ground state tuning (by pressure) to a SDW QCP [e.g. (TMTTF)$_2$X [116], (TMTSF)$_2$X [14], κ-(BEDT-TTF)$_2$X [120]], or even sometimes at ambient pressure [e.g., (TMTSF)$_2$ClO$_4$[121], κ-(BEDT-TTF)$_4$Hg$_{2.89}$Br$_8$ [40]].

As shown in Fig 5(a), two SC domes are observed in (TMTTF)$_2$SbF$_6$ with a QCP at the peak of the first SC dome,[116] consistent with other superconductors discussed so far. The resistivity exponent *n*, obtained by the same authors, reduces to ≤ 1 with decreasing doping, as depicted in the color-gradient plot in Fig. 5(a). The organic doped-material κ-(BEDT-TTF)$_4$Hg$_{2.89}$Br$_8$[40], which is a 'born' NFL material without tuning, also possesses two SC domes [see inset to Fig. 5(a)] as a function of pressure. Here the higher $T_c$ dome emerges at the lower pressure side; however, consistent with other systems the NFL region coincides with the higher $T_c$ dome. This counter example indicates that the higher-$T_c$ dome is not simply a manifestation of a higher value of a tuning parameter, *rather it appears where the parent ground state possess stronger NFL-behavior.*

## VB. Oxybismuthides

A new family of superconductors was discovered in 2012 namely, the oxybismuthides BaTi$_2$Pn$_2$O (*Pn* = As, Sb, and Bi) with $T_c \sim$ 1.2 - 5.5 K.[6], [119], [122]–[124] Here, superconductivity often arises in close proximity to a normal state density wave, which is thought to be either a SDW or a CDW, and likely unconventional.[124] In doped BaTi$_2$(Sb$_{1-x}$Bi$_x$)$_2$O, a SC dome was observed [118] at the QCP of the DW around x = 0.15, but the crossover from a FL to a NFL behavior occurs at a higher doping x ~ 0.3. With further increase in doping, superconductivity first disappears at x ~ 0.4. Following a non-SC region, up to x ~ 0.55 superconductivity reemerges, this time in a NFL phase and with a higher $T_c$ compared to the first SC dome.[119] To summarize the overall trend here, the two SC-domes phase diagram is reconstructed in Fig. 5(b) by combining data from Refs. [118], [119]. The result is essentially analogous to the phase diagram illustrated in Fig. 1.

## VC. Sr$_2$RuO$_4$

Sr$_2$RuO$_4$ is a distinct class of a 'born' superconductor, which is widely believed to possess chiral *p*-wave pairing symmetry.[125] Such a pairing symmetry is a natural choice in a FM background. This has led to many experimental investigations for the FM order and/or localized FM fluctuations in proximity to a SC state. Doping on both Sr and Ru sites with different elements yields different magnetic phase diagrams, however, superconductivity is sharply and monotonically suppressed in all cases without any sensitivity to magnetic phases. Ca$_{2-x}$Sr$_x$RuO$_4$ acquires a commensurate AF order, which transforms into a glassy (predominantly SG) phase with increasing doping, and appears to survive up to high doping.[126] The precise phase boundary between the SG and SC states is not yet well-established, however, the two phases appear to be well separated. In La$_{2-x}$Sr$_x$RuO$_4$, a NFL like state is seen to appear and optimized around *x* ~ 0.2.[127] The origin of the NFL state is argued to be associated with the FM fluctuations at the van-Hove singularity. In both materials, superconductivity arises only at *x* ~ 2.0, long after the AF order, SG phase and the NFL state have disappeared.

With isovalent substitution on the Ru site by Ti, i.e. in Sr$_2$Ru$_{1-y}$Ti$_y$O$_4$, superconductivity smoothly decreases by *y* ~ 0.2%.[126] With further doping, an incommensurate SDW phase (or a incommensurate SG phase) develops beyond the SC phase, and they clearly do not coexist.[126] An earlier study on this doped sample reported signatures of a NFL behavior in both the specific heat and resistivity; the resistivity exponent sharply decreased to 1 for *y* ~ 2.5%.[128] The SC regime (*y* = 0 - 0.2 %) is FL like and does not have a competing order. The authors argued that the NFL state in this doping side was triggered by 2D AF fluctuations due to FS nesting, rather than FM fluctuations as proposed for the NFL state in La$_{2-x}$Sr$_x$RuO$_4$.[126]

Magnetic doping by Co and Mn on the Ru site gives rise to a short-range FM order with Co doping, surviving up to x~2%, and an incommensurate AF or SG phase emerging with Mn doping.[129] In this phase diagram, the non-magnetic undoped sample at $x = 0$ may be considered to serve a QCP between the FM and AFM phases occurring with Co and Mn dopings, respectively. However, $x = 0$ sample exhibits simple FL like feature, and the system gradually becomes NFL like around 0.8% Co doping, and again becomes FM like by 1.5% Co doping. In all these samples, superconductivity does not coexist with any order phase, it is well-separated from the QCP and NFL state, and the latter two appear unrelated to each other.

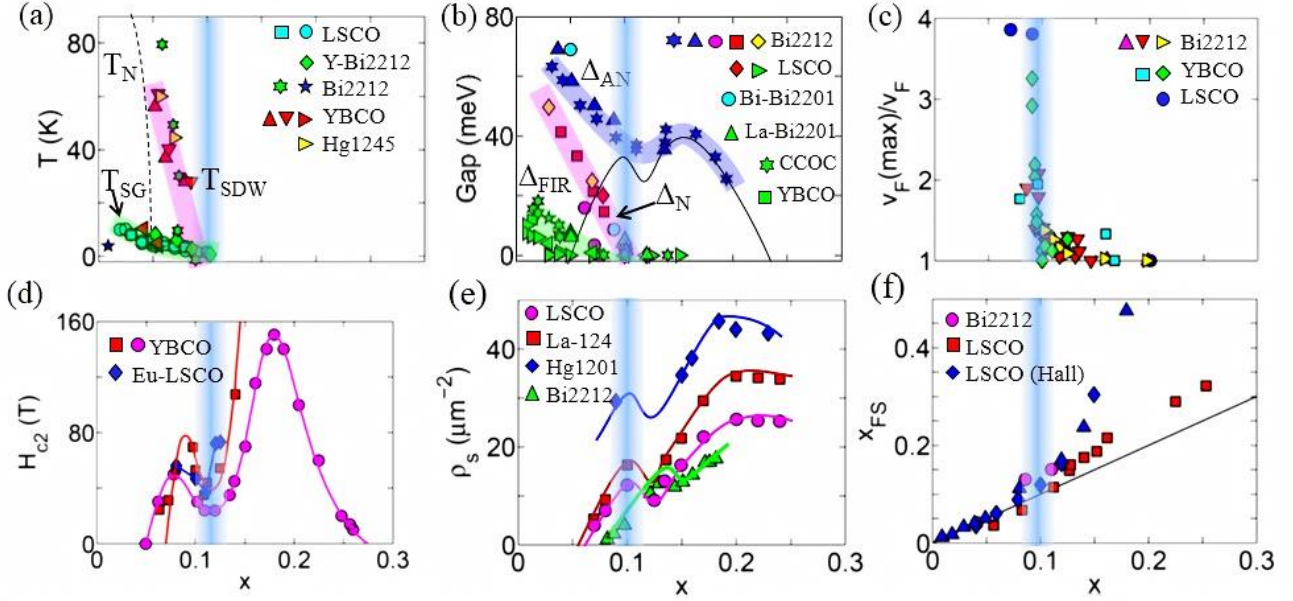

**Fig. 6.** Various properties of cuprates, exhibiting different anomalies at the two SC domes. (a) SG temperature ($T_{SG}$) for LSCO (circles are collection of data from Ref. [130], and squares from Ref. [83]), Y-Bi2212 [83], $Bi_2Sr_2CuO_{6+\delta}$ (Bi2201) [131], and YBCO (left triangle) [132]. SDW transition temperature ($T_{SDW}$) is shown for multilayer Bi2212[74], YBCO (upper triangle)[133], (lower triangles[134]), and multilayer Hg1245[75]. (b) Various low-energy gaps as observed in cuprates. Far-infrared optical gap ($\Delta_{FIR}$) collected for various hole doped systems.[135] A full gap is observed by ARPES in the SC state of underdoped samples, and attributed to a nodal gap ($\Delta_N$) for Bi2212 (magenta circles from Ref. [136] and red square from Ref. [137]) and LSCO[138]. The yellow diamond symbols, coinciding with the $\Delta_N$ values is obtained from FIR measurement[135]. The gaps at the antinodal point [$\sim(\pi,0)$], $\Delta_{AN}$, are measured by ARPES (star symbol from Ref. [136], [139] and triangles from Ref. [139]) and found to be consistent with transport results[140]. The cyan circles are the ARPES gap in Bi2201, which the authors attribute to a Coulomb gap.[141] (c) Inverse Fermi velocity (normalized to its maximum value near optimal doping) obtained from ARPES data on Bi2212 along the nodal direction[142], from SdH oscillation measurements on YBCO (diamond symbol),[143] and from thermal conductivity measurements for YBCO (square) and LSCO (filled circle).[144] (d) Upper critical field ($H_{c2}$) data are shown for YBCO[41] and Eu-substituted LSCO[86]. (e) Superfluid density at $T = 0$ is shown for LSCO,[83], [145] $HgBa_2CuO_{4+\delta}$ (Hg1201),[145] and Bi2212[146]. (f) FS volume versus hole doping per Cu site. The ARPES data for Bi2212, and LSCO are taken from the analysis of Ref. [147]. Blue diamonds depict LSCO data [148] from measurements of the Hall coefficient.

## VI. OTHER EVIDENCE OF DECOUPLED QCP AND NFL STATE

Next, we take advantage of the extensive literature available for cuprates and collectively analyze a wide range of experimental results where anomalies are observed either at the first SC dome and / or at the second SC dome (Fig. 6). We utilize available literature to identify properties which may be common among unconventional superconductors and therefore, assist towards the identification of key features. This may help elucidate the mechanism driving the NFL state in the SC dome with the higher $T_c$. Furthermore, this meticulous exercise provides a guideline for further, focused experiments across the wider range of unconventional superconductors. Numerous NMR, µSR and neutron experiments have reported evidence of a residual SG phase [83], [130]–[132], and / or a SDW [74], [75], [133], [134] extending up to a possible QCP at the first SC dome in several cuprates (Fig. 6(a)). These magnetic orders are connected to the underdoped Néel state and gradually become incommensurate at higher doping.[149] Interestingly, the temperature where the Nernst signal attains its maximum in various cuprates peaks around the first SC dome and not where $T_c$ is optimum i.e., in the second SC dome [150].

The presence of a corresponding magnetic gap in the electronic structure is also extensively documented by numerous spectroscopic investigations (Fig. 6(b)). High-resolution optical absorption spectroscopy revealed the presence of a gap in the far-infrared (FIR) region, which gradually disappears around the QCP of the possible SDW, i.e. at $x \sim 0.1$.[135] ARPES data in several cuprates provided evidence for a fully gapped electronic structure in the underdoped region and interpreted as disorder-induced Coulomb gap ($\Delta_C$),[141] or a nodeless SC gap ($\Delta_N$) [136], [137]. Finally, the quasiparticle gap in the antinodal region ($\Delta_{AN}$) follows the dome behavior of $T_c$ at higher dopings, while below $x \sim 0.1$ it starts to deviate from the same behavior of $T_c(x)$ and grows linearly with underdoping,[136], [139], [140] just like the magnetic gap.

Theories studying the coexistence of a density wave (DW) order and superconductivity may explain this doping dependence of $\Delta_{AN}$ as a combined gap of the two states [151], [152]. Extensions of the same theory predict that the coexistence of SDW and superconductivity leads to a third, triplet nodeless pairing symmetry[153]. This might explain the emergence of a fully gapped superconductivity (dictated by $\Delta_N$) below $x \sim 0.1$ where $\Delta_{AN}$ shows a kink. The three different gap structures, consistent with the doping dependence of $T_{SDW}$ and $T_{SG}$ suggest that a magnetic order extends up to $x \sim 0.1$ and coexists with superconductivity.

ARPES groups have reported an abrupt enhancement in the inverse Fermi velocity ($v_F$) around $x = 0.1$, see Fig. 6(c) [142]. Shubnikov de-Haas (SdH) oscillations also indicate that $v_F$ diverges around the same doping in YBCO [143]. Thermal conductivity measurements on cuprates[144] have shown that the bulk inverse Fermi velocity diverges at this doping (squares and circles in Fig. 6(c)). *The data suggest the possible presence of a QCP in cuprates at $x = 0.1$, i.e. the doping region where the first (lower $T_c$) SC dome has its maximum $T_c$.*

Next we examine the trends in the upper critical field ($H_{c2}$) and the superfluid density ($\rho_s$) [Figs. 6(d) and 6(e)]. The strong doping and field dependences of $H_{c2}$ in the two SC domes imply that both are unconventional. Furthermore, $H_{c2}$ and $\rho_s$ have two common parameters namely, the SC gap and the electronic effective mass as $H_{c2} \propto \xi_0^{-2} \propto \Delta^2/v_F^2 \propto \Delta^2 m^{*2}$ and $\rho_s \equiv \lambda_{ab}^{-2} \propto n_s/m^*$. Considering all the data together, it is apparent that superconductivity and magnetism coexist in the first dome, while the second SC dome perhaps escapes this mechanism.

Luttinger theorem can be used to test whether the NFL is manifested in the normal state FS. As pointed out by Gor'kov and Teitel'baum,[148] Phillips[147], by using the ARPES and Hall effect data, the Luttinger volume counting of the cuprate's FS is linearly proportional to the hole-doping per Cu atom only up to $x \sim 0.1$. At higher dopings the volume is larger than the actual doping concentration [Fig. 6(f)]. Hence, portions of the FS may become incoherent in the NFL state. Above the possible magnetic QCP, there is mounting evidence for the presence of AFM fluctuations,[83] paramagnons,[154] and charge fluctuations [155]. Detailed studies have further shown that some of the spin excitations move towards zero energy at the doping where the superfluid density is maximum[83]. These excitations may conspire to give rise to a NFL state[156] from which the second (high-$T_c$) dome might emerge as discussed in Sec. VIIA.

## VII. DISCUSSION AND OUTLOOK

In this section we converge our findings into the common properties present in all families of unconventional superconductors discussed here. For properties which tend to have multiple explanations and contradictory evidence (such as whether NFL and QCP are either related or decoupled from each other), we underscore the explanation that is universal. Our motif is to extract the minimum key parameters possibly responsible for the microscopic origin of superconductivity and the NFL behavior. Hence, we focus on the properties which are omnipresent in distinct families of unconventional superconductors, barring some materials or tuning parameters as exceptions due to lack of sufficient literature on relevant investigations. After pointing out these key properties, we address them in detail in terms of what existing models and experiments are consistent or inconsistent to explain the observed trends. Our aim is to encourage a focused, new research direction to advance the study of superconductivity including materials and tuning parameters.

Essentially, our study allows to obtain three key properties which emerge as universal across unconventional superconductors:

(1) *Two types of SC domes are present in all families of unconventional superconductors.* Superconductivity is unconventional in both domes, but with distinct normal and SC states properties. We categorize the SC regions as a lower-$T_c$ dome, which may or may not possess a QCP, and a higher-$T_c$ dome with a NFL or a strange metal normal state. These two domes are often merged into one, though can be separated when multiple tunings are employed simultaneously. In many materials the two domes are intrinsically separated along a single tuning parameter. In other materials, only one dome is present, depending on the nature of its normal state.

(2) *The NFL state and the QCP (if present) are often decoupled and well-separated.* This finding challenges the earlier belief that a QCP renders a NFL state in SC materials. In particular, we report extensive evidence indicating the presence of a large number of materials, which are 'born' NFL and SC. Suppressing the SC state in cuprates and pnictides using multiple tuning reveals no evidence for a QCP. Furthermore, engineering a QCP by doping with a magnetic atom yields either no change, or rather an opposite change in the resistivity exponent (reducing the NFL strength). Sometimes a QCP develops even at a value of the tuning parameter far from the NFL state (see Fig. 7 in Appendix A).

Also, we find in a large number of materials from all the families of unconventional superconductors studied here, that if there is a QCP along the phase diagram (far from the NFL state), the corresponding resistivity exponent is ~1.5, suggesting this state is not the dominant NFL. In some systems, although the QCP and the NFL state seemingly coincide, it is possible that the latter does not necessarily stem from a QCP.

(3) Returning to the corresponding SC properties in these two normal state phases, we find that the higher-$T_c$ dome arises at the NFL state, while the lower-$T_c$ dome is far from it. The lower $T_c$ dome may, or may not have a QCP. Since the NFL state shows linear resistivity, it often extrapolates to a lower residual resistivity coefficient [$\rho(0)$], implying that high $T_c$ emerges when the system possess a lower resistivity in the normal state[157]. A comparative study of resistivity exponent, residual resistivity and $T_c$ for various classes of superconductors is discussed in Appendix B. An important conclusion of the corresponding analysis is that *the NFL state appears to be an essential parameter for higher-$T_c$ superconductivity*.

**VIIA. Possible mechanism for the NFL state**

Our observations reveal that the resistivity exponent (*n*) is not minimum at the first SC dome, as one would expect for a possible QCP. On the other hand, *n* is minimum, reaching a value even below 1 at the second (higher-$T_c$) SC dome, emphasizing that this is a strange metal state, which seems to emerge in the absence of a clear QCP.

We discuss below the currently available explanations; focusing on their materials specific applications and consistencies, inconsistencies, and limitations in explaining the universal phase diagram identified in this study.

**1. Possible QCP in the higher-$T_c$ dome**

Due to experimental convenience, the possibility of a QCP in the second (higher-$T_c$) dome has been explored primarily in the family of cuprates, although some studies have been performed also in members of other families of unconventional superconductors. In particular, the possible QCP associated with the pseudogap has been suggested by experiments in several hole doped cuprates [79], [81]–[83], [158]. As mentioned earlier, the effective mass seems to diverge in Y123 at the second dome,[79] while it rather obtains a minimum here in Y124[80] indicating contradictory evidence for a QCP at the second dome within the same family. Similarly, as shown in Figs. 3(b) and 3(d) the extrapolation of the AFM and FM transition temperature in $CeNiGe_3$ and URhGe, respectively, is suggestive of a QCP at their second SC dome. In the H-doped La-1111 [87] and $KFe_2Se_2$ [39] families, the authors have discussed the possibility of an AFM critical point at the second SC dome. However, for a large number of other materials there is so far no evidence or indication for a distinct QCP in the higher-$T_c$ dome.

Notably, counter examples may argue against the possibility of a 'hidden' QCP beneath the second SC dome. For example, CeCoIn$_5$ and CeIrIn$_5$ are known as born NFL systems, in the absence of an apparent QCP. It has been suggested that a QCP is hidden nearby and can be accessed by doping. This system can be tuned to give three distinct behaviors [see Fig. 7]: (1) a QCP with magnetic doping on the In site, while with non-magnetic doping the QCP is removed. However, the NFL state remains unchanged in both cases; (2) with magnetic Rh ion doping on the Co/Ir site, a QCP emerges near 50% doping away from the NFL state – here, the NFL behavior at the QCP remains unchanged with and without the critical point; (3) by doping Yb, La and other actinides on the Ce site the NFL state shifts to a finite doping, with no evidence for a QCP. (It is noteworthy that a similar study can be performed on La-1111 and Ba-122 pnictide compounds since all sites can be doped and are known to have different phase diagrams.) The NFL state is a puzzle in unconventional superconductors and there have been systematic studies in the past three decades to understand its origin. Evidence for a QCP has been rather incomplete, materials specific and is lacking a systematic investigation. On the basis of experimental facts to date, we therefore conclude that only a NFL state appears to be ubiquitous for higher $T_c$ superconductivity. We also note that although the presence, or absence of a second QCP is not of direct relevance to the conclusions of the present work, a definite result would certainly allow for a better understanding of its role and implications with respect to the dominant NFL state in this regime of the phase diagram of unconventional superconductivity.

### 2. Density fluctuations

Varma et al [156] have shown that the spin and charge fluctuations without a dominant wavevector can give rise to a linear-in-energy self-energy, which in turn gives rise to a linear-in-$T$ resistivity. This particular form of the self-energy and the resulting anomalous phenomena are referred as marginal Fermi liquid. Generalization to other forms of density fluctuations is clearly possible. Spin and charge fluctuations are ubiquitous in all the unconventional superconductors studied here. In fact, early studies have shown the softening of short range spin-fluctuations at the second SC dome in cuprates[83]. Valence fluctuations are an inherent property of mixed-valence HF systems and have been discussed as a candidate mechanism for the second SC dome [9], [45], [53]. The importance of orbital fluctuations has been mentioned in the iron-pnictide families [159] and proposed as a possible mechanism for the second dome in La-1111 [37]. Nematic fluctuations surviving to the overdoped regime of pnictides is another candidate mechanism for the NFL state [160]. Counter arguments also exist where theories have predicted that quantum fluctuations do not always give rise to a resistivity linear in temperature [161], [162].

### 3. Disorder

The physics of disorder has been extensively studied in the context of semiconductors, correlated electrons as a possible reason for the NFL behavior in these materials.[7], [22], [23], [163], [164] In disordered systems, the random distribution of a microscopic parameter such as electron-electron interaction, Kondo exchange energy and density of states can lead to inhomogeneity and geometrical fluctuations. This in turn gives rise to properties such as phase separation and a metal-insulator transition. As the characteristic temperature separating the FL from the NFL is suppressed, the latter moves closer to the superconductivity dome. The physics of NFL due to disorder is extensively reviewed in Ref. [23] and in Refs. [7], [164]

It is however, worthwhile analyzing the materials where disorder driven NFL may be a possibility. Doped superconductors are possible candidates. However, as demonstrated in cuprates,[28], [32] heavy fermions,[53], [157] and pnictides[37], the application of multiple tuning via external disorder often leaves the NFL state unchanged. Moreover, the presence of a NFL at an optimum pressure or magnetic field makes a disorder-induced NFL an unlikely possibility.

### 4. Holographic NFL

More recently, the AdS/CFT framework has been employed to model strongly correlated condensed matter systems and correlated superconductors.[165] It is often seen that holographic entanglement has a gravity dual where long-range behavior leads to an anomalous resistivity-temperature exponent.[166]–[170] The exponent usually depends on various system parameters and can be tuned to 1 for a realistic range. This method is usually based on scale invariance and thus the applicability of holography may be argued as universal.

## VIIB. Electronic and magnetic orders

The role of electronic and magnetic orders for the formation of a SC dome and a QCP is a widely studied field. In most materials, electronic order in the first SC dome is rather well established, being AFM/SDW or FM (in U-based HF materials). The AFM/SDW (or FM) order seems to extrapolate to a QCP at the first SC dome. There is a possibility for a plethora of other competing orders, mainly in cuprates. Most studied orders include CDW, orbital order, a density wave, nematic order, and other exotic phases, including a pseudogap. Although in some cases these orders are present away from the first QCP, their role on the formation of a QCP at the NFL region is rare and materials specific. As discussed in the previous section, evidence across all families of unconventional superconductors suggest the NFL state in the higher $T_c$ SC dome can arise in the absence of a QCP. The purpose of this section is to address this observation in more detail.

### 1. Density waves

The AFM and/ or SDW ground states are perhaps the most common feature in unconventional superconductors. While AFM order often extends up to the first SC dome maximum, with a long coherence length, in some cases such as in hole-doped cuprates and in $FeSe_{1-x}Te_x$ it becomes a short-range SDW or a SG near the critical point. In other materials such as $CeFeAsO_{1-x}F_x$, it is also seen that AFM and superconductivity do not coexist, and superconductivity emerges precisely after the vanishing of the AFM order, both sharing the same QCP[94]. Such materials-dependent complexity is also observed in FM superconductors [see Fig. 4(d)]. On the other hand, a CDW is observed in some unconventional superconductors (we do not discuss the CDW state in transition metal dichalcogenides since superconductivity in these materials is often considered to be conventional). For example, in hole doped cuprates, a CDW arises around $x = 5\%$, where superconductivity also emerges. However, the CDW is optimal around 12-13% doping, that is, precisely where $T_c$ is suppressed between the two SC domes, suggesting a competing nature between the two orders. The CDW is projected to vanish around 18-20% doping, that is, near optimal superconductivity in the second SC dome, though further investigations are required to clarify this important possibility. Nevertheless, the apparently contradictory interplay between the CDW and the SC phases indicates either an indirect correlation between CDW and superconductivity, or the different role between charge order and the relevant extended fluctuations that might be present in the higher $T_c$ SC dome. Together with earlier studies on cuprates[83] this may suggest that spin and charge correlations play a fundamental role in the NFL phase of the higher $T_c$ SC dome of the phase diagram.

### 2. Pseudogap

A characteristic feature of hole-doped cuprates is the unsolved pseudogap state. Both the nature of the pseudogap (whether it originates from an electronic order, preformed pairs, or some form of fluctuations), and the precise location of its critical (end) point are not well established. In hole doped cuprates it has been argued that a QCP is present near the center of the second (higher-$T_c$) SC dome [79], [81]–[83], [158], or at the termination of the second SC dome [171], or even at other dopings. Evidence for a pseudogap has also been discussed in electron-doped cuprates[172] and the pnictide superconductors [173] although not as extensively as in hole doped cuprates. It is clear however, a pseudogap is not universal across all families of unconventional superconductors and therefore, is unlikely to be responsible for the apparently ubiquitous correlation between NFL and higher-$T_c$ superconductivity.

### 3. Nematic phase

The transition from the tetragonal lattice to the orthorhombic phase, as seen mainly in cuprates and pnictides, is often attributed to the onset of a $C_4$ symmetry breaking electronic order, known as nematic order [160], [174]. Doping dependence of the nematic phase is also highly materials specific; in cuprates the onset and critical behavior of the nematic phase is argued to coincide with the pseudoap temperature, the CDW, or the stripe phase [174]; while in the pnictides it is often seen above the SDW temperature and in many cases disappears with the suppression of SDW [160]. In a more recent study in $Ba(Fe_{1-x}Co_x)_2As_2$, it is shown that the nematic susceptibility has a QCP beneath the optimal SC dome.[101]

### 4. Other exotic orders and phase separation

Numerous other exotic electronic phases have been proposed for the various unconventional superconductors. Mesoscopic electronic orders related to the transition from a homogeneous to an inhomogeneous electronic structure such as stripe order, smectic phase, spin-charge separation, and others may be relevant to the separation of the two SC domes[174]. Evidence for this scenario can be obtained in hole-doped cuprates[175], $KSe_2Se_2$[176], and $CeNiGe_3$ [58].

### VIIC. Multiple tuning

Reconciling the diverse electronic, magnetic and SC properties emerging with different tuning parameters is an immense task and an active field of research. In general, tuning can lead to simultaneous or individual changes in the electronic structure, correlation properties, and/or pairing strength, and there is no general consensus on how a specific entity may be controlled. Many results in all the families of superconductors discussed here have demonstrated that $T_c$ is enhanced when the resistivity exponent ($n$) and the residual resistivity ($\rho_0$) are lowered, as discussed further in Appendix B. Therefore, we focus here on the available theoretical scenarios addressing how tuning may control the electronic orders and the NFL state, and thereby contribute in the formation of the SC dome(s).

In general, a SC dome develops when Mott insulators (cuprates) or mixed valence HF's, or correlated semimetals (pnictides, chalcogenides) are tuned to their metallic ground state. The parent compounds are dominated by localized electrons loosing sufficient Fermi energy required for superconductivity. As the systems are driven towards a metallic regime by adding carriers, applying pressure, etc., the residual localized and conducting electrons coexist and compete. Due to combined effects dominating the intermediate coupling regime, instabilities may develop including density-density fluctuations, mesoscopic inhomogeneities, phase separation and critical phenomena. These properties have a tendency to suppress phase coherence and quasiparticle lifetime, leading to NFL-like states. Tuning can control the spectral weight distribution between the localized and itinerant states, and provide the optimal correlation strength for many phases, which include a NFL state and a high-$T_c$. Notably, this scenario has been addressed in various forms in the context of cuprates [13], [155], [177], pnictides [178], [179], as well as in HF materials [9], [180].

Superconductivity is enhanced both around the QCP and the NFL state. Inherently, two SC domes may emerge separately, individually or even merged into a single dome. The role of complementary tuningis twofold: it helps reveal the two underlying SC domes, *and enhances the strength of the NFL state and/or pairing potential*. The latter possibility has been demonstrated in several SC families. For example, in cuprates, as shown in Fig. 2(d), the optimal $T_c$ in Bi2223 is enhanced from about 110 K to above 140 K with pressure[34]. In doped $CeRh_{1-x}Ir_xIn_5$, superconductivity increases with pressure, as shown in Fig. 4(c)[35]. In $KFe_2Se_2$, superconductivity is optimal at the Fe vacancy level of around 1.6. As shown in Fig. 5(c), pressure gives rise to two SC domes in this sample, with superconductivity enhanced in the second dome.

### VIID. Outlook

The present study indicates a tantalizing link between NFL physics and higher $T_c$ superconductivity. It calls for further studies across unconventional superconductors to establish a more quantitative link and encourage the development of a theoretical understanding as well as the prediction of new materials. From the context of this article, further research activities can be pursued in several directions. Considering the importance of multiple tuning for unraveling fundamental normal and SC properties, this is certainly an appealing research direction to pursue. In cuprates, multiple tuning parameters include disorder, pressure and magnetic field. Disorder dependence has been extensively studied in a number of materials, unravelling two SC domes. However, pressure and magnetic field dependencies have been studied primarily in Bi2223 and YBCO, respectively, and can be explored further. Moreover, the evolution of the resistivity exponent as a function of these two latter parameters is not well known. Finally, the study of multiple tuning is lacking in Hg-, Tl-, and Cl-based curates, all of which exhibit characteristic intertwined orders and FS properties.

Magnetic field dependence on pnictides has not been explored as extensively as in the HF's and cuprates. As mentioned in Sec. IVA, F-doped La-1111, Ce-1111, and Sm-1111 have characteristically different phase diagrams. Therefore, additional doping on the La site with Ce or Sm and vice versa can be a fertile ground to investigate the competition between SDW and superconductivity. Much like how simultaneous dopings at different sites of SmFeAs$_{1-y}$P$_y$O$_{1-x}$H$_x$ unraveled two SC domes [Fig. 5(a)], it may be instructive to perform similar studies in the (Ba$_{1-x}$K$_x$)(Fe$_{1-y}$Co$_y$)$_2$(As$_{1-z}$P$_z$)$_2$ ('122') family. Finally, LiFeAs [181] and NaFeAs[182] are members of the '111' pnictide family, which have SC domes with and without a SDW QCP, respectively. It would be interesting to grow doped Li$_{1-x}$Na$_x$FeAs to study the evolution of the two SC domes.

HF compounds are often tuned with pressure or magnetic field. Disorder studies have been carried out only on a few compounds. Similar to the above discussions, the Ce-115 family Ce$_x$$R_{1-x}$$M_{1-y}$$M_y$'(In$_{1-z}$$T_z$)$_5$ ($R$= Yb, La; $M$, $M'$=Co, Rh, Ir; $T$=Hg, Cd, Sn, Pt) (discussed in Appendix A) provides a versatile platform to study via multiple dopings the presence of a NFL, with and without a QCP. In other families of HF's, e.g. Ce$R_2$$M_2$ ($R$ = Cu, Pd, and Ni; $M$ = Si, Ge), the interplay between AFM, valence fluctuations and Kondo physics, and their role on the NFL state can be uniquely explored via multiple tuning. Furthermore, U-based compounds have different phase diagrams for different materials, as shown in Fig. 4(d). The possibility of doping UGe$_{2-x}$Rh$_x$ or UGe$_{2-x}$Co$_x$ and the application of pressure provide an opportunity to monitor the FM QCP and the NFL state, as well as the possible role of FM fluctuations on the NFL. Finally, the Pu-115 family [such as Pu$MR_5$ ($M$ = Co, Rh; $R$ = Ga, In)] has been realized recently to be SC with highest $T_c$ ~ 18.5 K [183]. Although these materials are believed to be unconventional, there is no compelling evidence to date for an electronic order or a QCP, and the NFL physics has yet to be explored.

Strain is another promising tuning parameter for unconventional superconductors. Strain-dependent studies in pnictides [104] and HF compounds[56] find significant enhancement of superconductivity. Uniaxial strain dependence can be an important tool to pin down the role of nematic fluctuations in the NFL state and superconductivity.

With the rapidly increasing availability of high-quality bulk single crystals, atomic layer deposition techniques and high-resolution spectroscopies, we are in an exciting period for superconductivity. High-precision measurements in the SC state are now possible in many families of unconventional superconductors. Therefore, much like in cuprates, we hope our study would motivate further investigations through multiple tuning also in other families of unconventional superconductors, especially in iron-pnictides and HF's where the sample variety and quality continue to be improved.

### Acknowledgements.

We are indebted to many colleagues for useful discussions, in particular, to L. Greene, A. Leggett, G. Lonzarich, B. Maple, R. Markiewicz, D. Pines, T.V. Ramakrishnan, J. Schmalian, J.-H. She, A. Soumyanarayan, F. Steglich, and J. Thompson. CP acknowledges financial support from the National Research Foundation, Singapore, through the National Research Foundation (NRF), NRF-Investigatorship (NRFNRFI2015-04). TD acknowledges financial support from the Science & Engineering Research Board (SERB) on the category of Start Up Research Grant (Young Scientist).

# APPENDICES

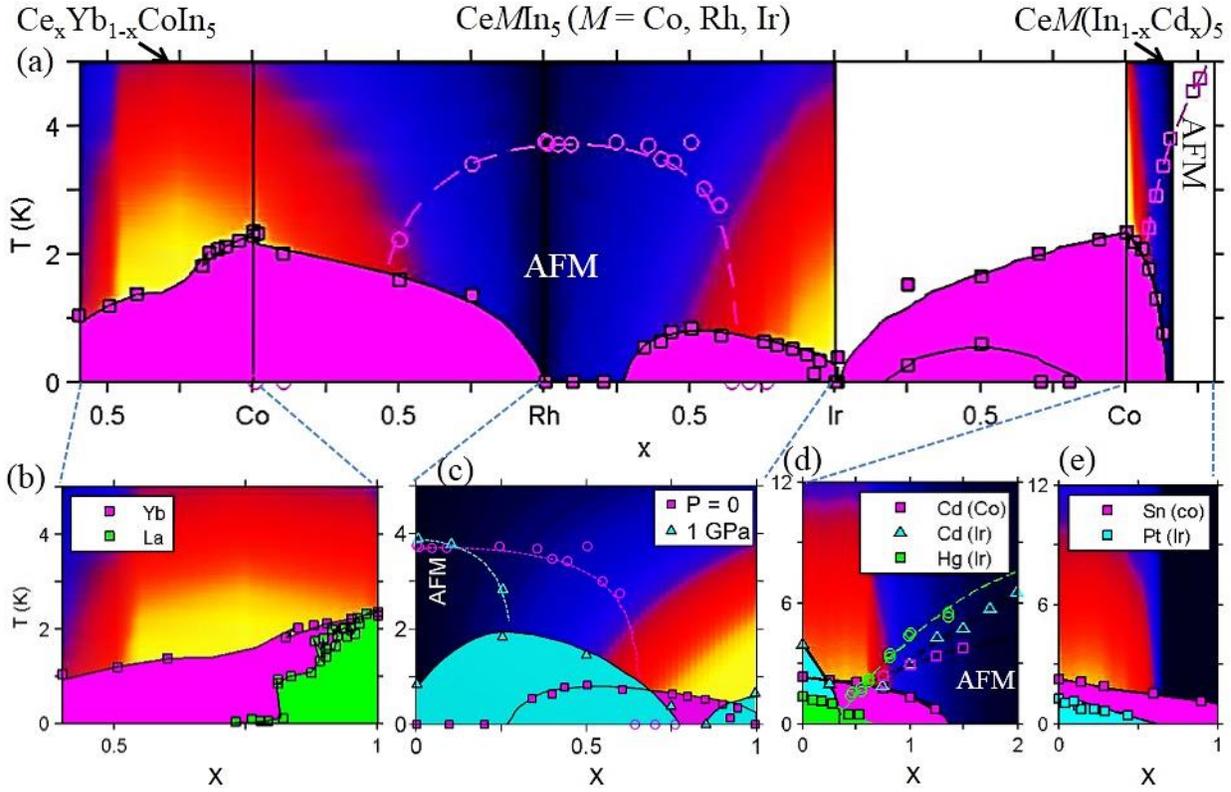

**Fig 7.** Evolution of the phase diagram as a function of multiple tuning at different sites of CeCoIn$_5$. (a) The AFM and SC phase diagram for Co$M$In$_5$ ($M$ = Co, Rh, and Ir) is taken from Ref. [48]. The data for doping on Co and In sites are added here. (b) The data for Ce$_x$R$_{1-x}$CoIn$_5$ is taken from Refs. [45], [184] (c) The pressure dependence data for CeCo$_{1-x}$Rh$_x$In$_5$ is obtained from Ref. [35]. (d) The Cd and Sn dopings data are taken from Ref. [49] and Ref. [185]. (e) Data for CeIr(In$_{1-x}$T$_x$)$_5$ ($T$ = Hg and Pt) are obtained from Ref.[50]. No magnetic phase is observed for Sn and Pt dopings.

## A. More heavy fermions: Ce$_x$R$_{1-x}$M$_{1-y}$M'$_y$(In$_{1-z}$T$_z$)$_5$ ($R$= Yb, La; $M$, $M'$=Co, Rh, Ir; $T$=Hg, Cd, Sn, Pt )

As mentioned in Sec. IIIB, the HF Ce-115 family can be tuned via multiple methods, including doping at every ionic site, pressure and magnetic field. Data obtained using various possible dopings is compiled in Fig. 7. In CeCo$_{1-x}$Rh$_x$In$_5$, the doping versus $T_c$ plot does not exhibit an observable anomaly at the QCP, while an enhancement in $T_c$ would be expected here. On the other hand, CeRh$_{1-x}$Ir$_x$In$_5$ exhibits a SC dome around a QCP[47]. However, in both cases the NFL state remains indifferent to the position of the QCP (~50% doping), with the NFL state remaining dominant in the undoped case. As discussed in the main text, the application of pressure helps split the SC dome into two, and with increasing pressure the optimum $T_c$ in both domes increases.[35]

The In-site of Ce-115 has been doped with Cd (Ref. [49]) and Hg (Ref. [50]) giving rise to a magnetic phase. Similarly, doping with Sn (Ref. [185]) and Pt (Ref. [50]) also give rise to a magnetic phase. In magnetic CeRh(In$_{1-x}$Cd$_x$)$_5$, Cd doping first suppresses the AFM Nèel temperature ($T_N$) and with 7-8% Cd doping $T_N$ shows a sudden upturn. No sign of superconductivity has been reported in this system. In CeCo(In$_{1-x}$Cd$_x$)$_5$, superconductivity is suppressed while $T_N$ increases with doping and extrapolation of the $T_N$ curve points to the possible presence of a QCP around 2-4% doping. On the other hand, in CeIr(In$_{1-x}$Cd$_x$)$_5$ superconductivity is suppressed rapidly and with 2-3% doping AFM emerges. The phase diagram of CeIr(In$_{1-x}$Hg$_x$)$_5$ is however, similar to CeCo(In$_{1-x}$Cd$_x$)$_5$. Sn and Pt doping gradually suppress superconductivity with no evidence for an AFM state. In both doping cases, the resistivity exponent increases monotonically from the pristine material and the systems gradually loose the NFL signature, irrespective of the presence or absence of a QCP.

Rear earth substitution on the Ce site in this family has been extensively studied by Maple's group [45], [184] to explore the role of valence fluctuations on superconductivity. Yb doping suppresses superconductivity without a magnetic state arising in CeCoIn$_5$. Interestingly, the resistivity exponent ($n$) decreases even further, from the value of ~1 in the pristine sample to as low as 0.3-0.4, or even lower for ~40% Yb doping, [184] and then increases again. No evidence for a QCP is found for dopings as high as 60%.

## B. Residual resistivity

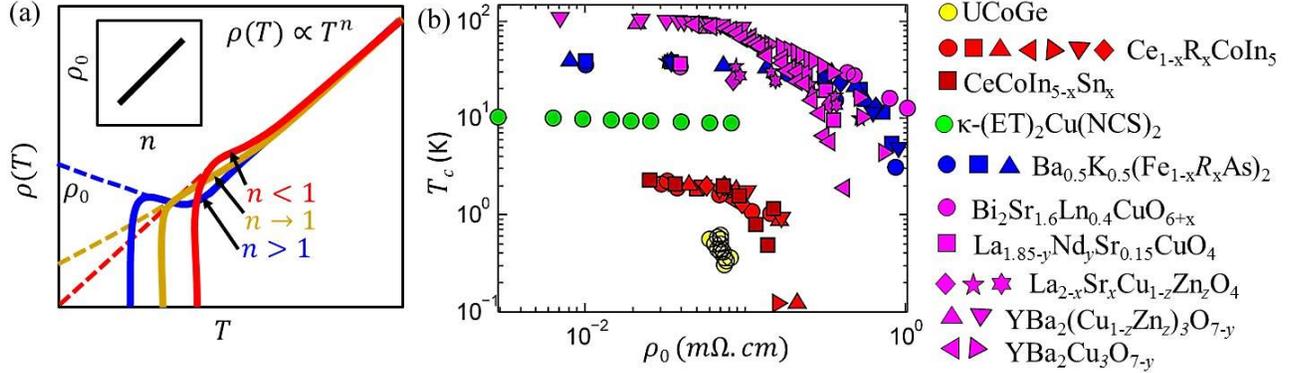

**Fig. 8:** Residual resistivity versus resistivity exponent and $T_c$. (a) Schematic drawing of the temperature dependence of resistivity in unconventional superconductors (normalized at high temperature) for three representative cases. This figure illustrates how the residual resistivity is expected to be reduced as the corresponding exponent decreases (see inset). (b) As deduced for most of the systems studied in the main text, higher-$T_c$ is associated with a lower resistivity exponent, hence, lower residual resistivity. This observation is supported by numerous results in heavy-fermion [157], [185], [186], organic [187], pnictide [188], and cuprate [189]–[191] superconductors.

The resistivity-temperature exponent ($n$) is shown in the main text to give vital clues towards the route to higher-$T_c$, since the optimum $T_c$ in most of the materials studied here is located where $n$ is minimum. An important consequence of the resistivity exponent is that the smaller its value the smaller the residual resistivity. To illustrate this feature, we take three representative cases in Fig. 8(a) for $n > 1$ ($n = 1.5$-$2$ being a FL state), $n = 1$, and $n < 1$ (both are considered NFL). As $n \to 2$, the projected resistivity at $T = 0$ increases (blue lines), and as $n$ decreases the corresponding residual resistivity $\rho_0$ decreases. Interestingly, $T_c$ also decreases with increasing $n$, i.e., with increasing $\rho_0$.

This fact is already known and studied both theoretically and experimentally in various families. We present $\rho_0$ versus $T_c$ data from the literature [Fig. 8(b)]. Results are shown for various representative materials such as the FM superconductor UCoGe,[186] HF superconductors without a QCP in Ce$_{1-x}$R$_x$CoIn$_5$ ($R$ = rear earth elements),[157] and Sn doped CeCoIn$_{5-x}$Sn$_x$ [185]. The $\rho_0$ versus $T_c$ data appear flat in Fig. 8(b) due to large y-axis range, but the actual data in Ref. [187] indicate that $T_c$ increases with decreasing $\rho_0$. The pnictide data are shown for an SDW superconductor in the 122 family for three different impurities ($R$ = Mn, Cu, and Ni) [188] on the Fe site, depicting the same behavior. Within the cuprate family, data are shown for a variety of different materials including $Ln$-Bi2201 ($Ln$ = rear earth),[189] and Nd-LSCO,[189] Zn-LSCO,[190] Zn-YBCO and Nd-YBCO, [190] and YBCO [191].

Different materials have different impurity levels and other factors that control the strength of superconductivity, hence the diversity in the curves. However, in all these families, the higher $T_c$ is obtained when the corresponding $\rho_0$ for that material is reduced. This result demonstrates that the route to higher-$T_c$ may be a mechanism able to reduce the normal state resistivity exponent and the residual resistivity in the NFL state.

**Table 1**: Separate list of materials having one SC dome and two SC domes. Those materials exhibiting one dome, have either a QCP (without a strong NFL behavior, i.e., the resistivity exponent at the QCP is ~1.5), or a NFL state with any apparent QCP, or a coexistence of the two. Also in some of these materials, multiple tuning has not been applied yet and therefore cannot dissect the SC dome. The number of materials exhibiting two SC domes is significantly higher than the ones with one dome.

| One SC dome: QCP, NFL states *coincide* or, either a QCP or a NFL state is only present. | | |
|---|---|---|
| **Family** | **Materials** | **Remarks** |
| Heavy fermions | $CeIn_3$ | Minimum value of $n \to 1.6$ at the AFM QCP. No second dome has been observed.[192] |
| | $CePd_2Si_2$, | $n \to 1.3$ at the AFM QCP. No second dome has been observed.[56] |
| | $CePt_2In_7$ | Superconductivity has an extended dome, as in many pnictides, with an AFM QCP intercepting the lower part of the SC dome.[193] |
| | $YbRh_2Si_2$ | AFM QCP & NFL coincide, but the *SC state vanishes before the QCP/NFL point*.[194] |
| | UCoGe | SC dome, FM QCP, and NFL state ($n \to 1$) coincide.[61] However, with Si doping on Ge, the *SC state above the QCP is completely suppressed and it survives only within the FM state*.[59] A similar phase diagram is observed for $UGe_2$,[51] $ZrZn_2$.[52, p. 2] |
| Cuprates | Electron doped systems | AFM QCP and NFL coincide around 15%-16% doping.[172] Evidence for a second SC dome is rare here.[84], [85] |
| Pnictides | '1111' Family: Only $SmFeAsO_{1-x}F_x$ | Among the '1111' family, only in this system the AFM QCP penetrates into the SC dome.[96] NFL is close to the QCP point, but not at the same doping. |
| | '122' Family: Only $Ba(Fe_{1-x}Co_x)_2As_2$, and $BaFe_2(As_{1-x}P_x)_2$ | $Ba(Fe_{1-x}Co_x)_2As_2$: SDW QCP and SC dome coincide at the same doping.[14] Recently, a nematic QCP is detected at the optimal doping of this compound.[101] In contrast, the hole doping side has a considerably wider SC dome, extending beyond the QCP. The NFL state is not dominant at the QCP (see below).[195] $BaFe_2(As_{1-x}P_x)_2$: AFM QCP coincides with NFL. Mass enhancement and superfluid density suppression are cited as evidence for the QCP.[15], [99] |
| Organics | $(TMTSF)_2PF_6$ | SDW QCP coincides with the optimum SC dome. $n \to 1$ near the QCP.[14] |
| Ruthenate | $R_{2-x}Sr_xRu4$ ($R$ = Ca, La) and $Sr_2Ru_{1-y}M_yO4$ ($M$ = Ti, Co, Mn) | Superconductivity does not possess a characteristic dome shape and sharply disappears with all dopants. Sometimes a FM or AF order develops above the SC phase, and a NFL state is detected in a few doped samples. However, neither an order, nor the QCP, or the NFL state coexist with the SC phase.[126], [127], [127]–[129] |
| **Two SC domes: NFL state arises either at a *different* location or without a QCP** | | |
| Heavy fermions | 'Ce-122' family: $CeCu_2Si_2$, $CeRu_2Si_2$, $CePd_2Si_2$, $CeCu_2Ge_2$ | $CeCu_2(Si_{1-x}Ge_x)$: Two separate SC domes are observed. $n \to 1.5$ at QCP (under the low-$T_c$ dome), and $n \to 1$, at the high-$T_c$ dome without a QCP.[30] $CeRu_2(Si_{1-x}Ge_x)_2$: *No* mass enhancement occurs at the QCP.[55] $CePd_2Si_2$: AFM QCP, NFL ($n \to 1.3$) and superconductivity have the same strain dependence, with |
| | | no anomaly at the QCP.[56] |
| | CeNi$_2$Ge$_2$, CeNiGe$_3$ | Both systems show two SC domes, with at least one dome inside the AFM state.[57], [58] There is no clear QCP at the NFL state. Homogenous to inhomogenous phase transition between the two SC domes is argued in CeNiGe$_3$, as in hole-doped cuprates. |
| | '115' family: Ce$_x$R$_{1-x}$M$_{1-y}$M'$_y$(In$_{1-z}$T$_z$)$_5$ (R= Yb, La; M, M'=Co, Rh, Ir; T=Hg, Cd, Sn, Pt ) | Doping on the *In* site: <br> (a) With magnetic doping (Cd, Hg), a QCP is seen.[49], [50] <br> (b) With, non-magnetic doping (Sn, Pt), no QCP is seen while the NFL state stays intact.[50], [185] <br> Doping on the *Co/Ir* site: With magnetic Rh, a QCP is engineered however, at around 50% doping away from the location of the NFL state.[35], [47] <br> Doping on the *Ce* site: With various actinide substitutions on the Ce site, the NFL state shifts to higher doping, with *no* evidence for a QCP.[45], [184] |
| | URhGe, | Two SC domes emerge with in-plane magnetic field, without a prominent FM QCP.[42] |
| | UGe$_2$, ZrZn$_2$, UCoGe$_{1-x}$Si$_x$ | Although these materials appear to have one SC dome, the FM QCP and superconductivity disappear at the same doping/pressure, where NFL is dominant.[51], [52], [59] |
| | U$_{1-x}$M$_x$Pd$_2$Al$_3$ (M = Th, Y, La), M$_{1-x}$U$_x$Pd$_3$ (M = Sc, Y) | In both systems,[64]–[66] the NFL state arises above the QCP, often intervened by a SG phase. No evidence for a second SC dome. |
| | URu$_{2-x}$Re$_x$Si$_2$, Yb$_2$Fe$_{12}$P$_7$ | The NFL phase is observed deep inside the magnetic phase, not at the FM QCP.[67], [69] No evidence for a second SC dome. |
| Cuprates | Hole doped systems (La-family, YBCO, Bi-family, Hg-family) | QCP in first dome: In *some* cuprates, a SG or residual a SDW (not the parent AFM state) has a QCP at the first SC dome with $n \geq 1.5$ (see Fig. 6(a)). <br> QCP at second dome: $n \leq 1$ here. In YBCO, and some Bi-based compounds, a CDW is recently observed below optimal doping whose QCP is not yet unveiled.[155] The pseudogap state is materials specific and no clear indication of a QCP at the NFL state is presented here except in YBCO. Only in the latter system, evidence for amass enhancement at the second (higher-$T_c$) SC dome is presented.[79] Evidence for the persistence of low energy magnetic correlations in the second dome and up to optimal superconductivity suggest the possible role of the former to the presence of a robust NFL[84]. <br> Stripes: Homogeneous to inhomogeneous phase separation at the 1/8 doping is evidenced in cuprates[12], [174]; the strength decreases gradually from LBCO, to LSCO, Bi2212, YBCO, and HBCO. |
| | Electron doped systems (Sr$_{1-x}$La$_x$CuO$_2$, La$_{2-x}$Ce$_x$CuO$_4$) | Sr$_{1-x}$La$_x$CuO$_2$ thin film grown on DyScO$_3$ shows two connected SC domes as a function of doping.[84] <br> La$_{2-x}$Ce$_x$CuO$_4$: The presence of a long tail above the optimal doping in this system hints for the presence of a second dome which may be revealed with multiple tuning.[85] |
| Pnictides | '1111' family: LaFeAsO$_{1-x}$F$_x$, CeFeAsO$_{1-x}$F$_x$, | LaFeAsO$_{1-x}$F$_x$: First order SDW phase transition, and SC appear after SDW.[95] More recently, two SC domes have |
| | $Ln$FeAs$_{1-y}$P$_y$O$_{1-x}$H$_x$ ($Ln$ = La, Ce, Sm, and Gd) | been observed in this compound with QCP and NFL states separated in different domes.<br>CeFeAsO$_{1-x}$F$_x$: A SC state arises only after SDW is suppressed .[94]<br>$Ln$FeAs$_{1-y}$P$_y$O$_{1-x}$H$_x$: Two SC domes emerge with a SDW QCP in the first dome ($n \geq 2$) and a NFL state ($n \leq 1$) in the second dome, without a QCP.[88] |
| | '122' family:<br>Ba$_{1-x}$K$_x$Fe$_2$As$_2$,<br>$M$Fe$_2$As$_2$ ($M$=K, Cs, Rb),<br>KFe$_2$Se$_2$,<br>$R$Ni$_2$Se$_2$ ($R$=Ti, Ba, Sr) | Ba$_{1-x}$K$_x$Fe$_2$As$_2$: The two connected SC phases are nodeless and nodal in character, and present at the two extreme doping limits, respectively. The optimal doping is about 20% higher than the SDW QCP in this system. Interestingly, $n \geq 1.5$ at the QCP, and decreases to $n \leq 1$ around optimal doping, without an apparent QCP.[195]<br>$M$Fe$_2$As$_2$: A strong suppression of $T_c$ at an intermediate pressure is observed in all materials in this family without a clear indication for a QCP.[38], [89], [90]<br>KFe$_2$Se$_2$: Two distinct SC domes as a function of pressure, with no clear indication for a QCP.[39]<br>TiNi$_2$Se$_{2-x}$S$_x$: One SC dome at $x$=2, splits into two at $x$=0.[113] |
| | '11': Fe(Se,Te) | $\alpha$-FeSe$_{0.88}$ and $\beta$-Fe$_{1.01}$Se show two SC domes. Little information about a QCP and a NFL is available.[91], [92] |
| | '111' family:<br>LiFeAs,<br>NaFeAs | LiFe$_{1-x}$Co$_x$As: Superconductivity decreases monotonically in the absence of magnetism.[196]<br>NaFe$_{1-x}$Co$_x$As: Magnetism is very weak and disappears very quickly, yet superconductivity is as strong as in other pnictide families.[196] |
| Organics | $\beta$-(BEDT-TTF)$_4$Hg$_{2.89}$Br$_8$ | Two SC domes are seen, with $n \rightarrow 1$ at the high-$T_c$ dome, and $n \rightarrow 2$ at the lower-$T_c$ dome. No evidence for a QCP in either dome.[40] |
| | (TMTTF)$_2$SbF$_6$ | Two SC domes with one at a QCP and the other around a NFL.[116], [117] |
| Oxybismuthides | BaTi$_2$(Sb$_{1-x}$Bi$_x$)$_2$O | Two fully isolated SC domes: lower-$T_c$ dome at a QCP with $n \rightarrow 2$, and the higher-$T_c$ dome is NFL ($n \rightarrow 1$) without a QCP.[118], [119] |